\documentclass[10pt,a4paper]{article}

\usepackage[utf8]{inputenc}
\usepackage[margin=1in]{geometry}
\usepackage{amsmath}
\usepackage{amssymb}
\usepackage{graphicx}
\usepackage{booktabs}
\usepackage{cite}
\usepackage{hyperref}

\title{\textbf{Tensorized Radiative Heat Transfer for a Scalable and Calibrated Building Energy Simulator}}

\author{
  \textbf{Sang woo Ham}\thanks{Lawrence Berkeley National Laboratory, Berkeley, CA, USA. Email: \texttt{sham@lbl.gov}} \and
  \textbf{Donghun Kim}\thanks{Lawrence Berkeley National Laboratory, Berkeley, CA, USA. Email: \texttt{donghunkim@lbl.gov}} \and
  \textbf{Michael Rossetti}\thanks{Google, Georgetown University, New York University, George Washington University, USA. Email: \texttt{mjrossetti@google.com}} \and
  \textbf{John Sipple}\thanks{Google, USA. Email: \texttt{sipple@google.com}}
}

\date{}

\begin{document}

\maketitle

\begin{abstract}
Accurate building energy simulation is essential for developing advanced control strategies that enable demand flexibility and grid responsiveness. The Smart Buildings Control Suite (\texttt{sbsim}) offers a lightweight, scalable, and data-calibrated simulation environment based on a tensorized finite-difference model. Previous work extended \texttt{sbsim} to include interior long-wave radiative heat exchange between indoor surfaces. However, a complete thermal model must also account for exterior radiative processes, including long-wave radiation exchange with the sky and surroundings, as well as short-wave solar radiation incident on building surfaces. This paper presents a comprehensive radiative heat transfer implementation for \texttt{sbsim} that integrates both interior and exterior radiation mechanisms. Our primary contribution is the development and integration of a fully tensorized exterior radiation module that captures sky and ground long-wave exchange as well as solar heat gains through opaque and transparent surfaces. By formulating these processes as tensor operations compatible with the existing framework, we preserve the computational efficiency necessary for reinforcement learning applications. We validate our implementation against established simulation tools and demonstrate improved prediction accuracy for surface temperatures and building thermal loads. This enhancement significantly increases the physical fidelity of \texttt{sbsim}, enabling more realistic training environments for building energy optimization and control.
\end{abstract}

\section{Introduction}

The U.S. electrical grid confronts mounting pressure from the rapid expansion of data-intensive industries, underscoring the need for flexible demand-side resources \cite{NERC2024-by,NERC2024-lr}. Buildings consume approximately 74\% of the nation's electricity \cite{US-EIA2025-lc} and represent a substantial opportunity for load flexibility. HVAC systems, as thermostatically controlled loads, can exploit a building's thermal inertia to shift consumption away from peak periods \cite{Neukomm2019-ze}. Reinforcement learning (RL) has emerged as a promising paradigm for orchestrating such flexibility \cite{Touzani2021-su}, but training agents directly in occupied buildings risks discomfort and equipment damage, making computationally efficient simulators indispensable \cite{Goldfeder2024-qb,Goldfeder2023-iq}.

The Smart Buildings Control Suite (\texttt{sbsim}) was designed to fill this role \cite{Goldfeder2024-qb}. Its core is a two-dimensional finite-difference thermal model expressed as tensor operations, enabling GPU-accelerated execution practical for RL training. The initial release captured conduction and convection but omitted radiation. A subsequent extension added interior long-wave (LW) radiation exchange among indoor surfaces using a fully tensorized view-factor approach \cite{Ham2025-ht}, demonstrating that radiative physics could be incorporated without sacrificing computational speed.

A physically complete model must also represent exterior radiative processes. Exterior surfaces exchange LW radiation with the sky, ground, and atmosphere---phenomena that influence nighttime cooling and surface temperatures. Short-wave solar radiation absorbed by opaque surfaces and transmitted through glazing constitutes a dominant daytime heat gain. Omitting these mechanisms limits predictive accuracy and may yield control policies that transfer poorly to real buildings.

This paper extends \texttt{sbsim} with a comprehensive radiative module. Our contributions are: (1)~a tensorized exterior LW radiation model parameterized by tilt angle and emissivity; (2)~a solar radiation model distinguishing absorbed and transmitted components, with transmitted radiation coupled to interior thermal mass; (3)~interior mass nodes capturing the thermal lag of furnishings and floor slabs; and (4)~numerical validation against an iterative solver demonstrating a 4.19$\times$ speedup.

\section{Background}

The \texttt{sbsim} framework discretizes each building floor into a two-dimensional grid of control volumes (CVs) and advances the temperature field through tensor arithmetic. An energy balance is formulated for every CV, accounting for conduction between neighboring volumes, convective exchange with ambient or zone air, and any externally applied heat sources. The update equation computes the new temperature tensor $T$ as \cite{Goldfeder2024-qb}:
\begin{equation}
\label{eq:original_sbsim}
\begin{aligned}[t]
T ={} & \biggl[ Q_x + Vz \bigl( K_1 U^{-1} T_1 + H_1 T_{\infty} + K_3 U^{-1} T_3 + H_3 T_{\infty} \bigr) \\
      & + Uz \bigl( K_2 V^{-1} T_2 + H_2 T_{\infty} + K_4 V^{-1} T_4 + H_4 T_{\infty} \bigr) \\
      & + \tfrac{C\rho UVz}{\Delta t}\, T^{(-)} \biggr] \\
      & \cdot \biggl[ Vz \bigl( K_1 U^{-1} + H_1 + K_3 U^{-1} + H_3 \bigr) \\
      & + Uz \bigl( K_2 V^{-1} + H_2 + K_4 V^{-1} + H_4 \bigr) + \tfrac{C\rho UVz}{\Delta t} \biggr]^{-1}
\end{aligned}
\end{equation}
where $Q_x$ is an external heat source (e.g., HVAC), $K$ and $H$ denote conductivity and convection coefficient tensors for each cardinal direction, $U$ and $V$ are CV dimension tensors, $C$ and $\rho$ are specific heat and density, $z$ is the floor height, and $T_1$--$T_4$ are spatially shifted temperature fields. All operations are element-wise, and the equation is iterated until the maximum temperature change across all CVs falls below a convergence threshold. This formulation captures conduction and convection but contains no radiative terms. Interior LW radiation was subsequently added through a pre-computed view-factor matrix that encodes geometry and emissivity information; the reader is referred to \cite{Ham2025-ht} for a complete description of the view-factor calculation and its integration into the tensor framework.

\section{Methodology}

\begin{figure}[htbp]
  \centering
  \includegraphics[width=0.7\linewidth]{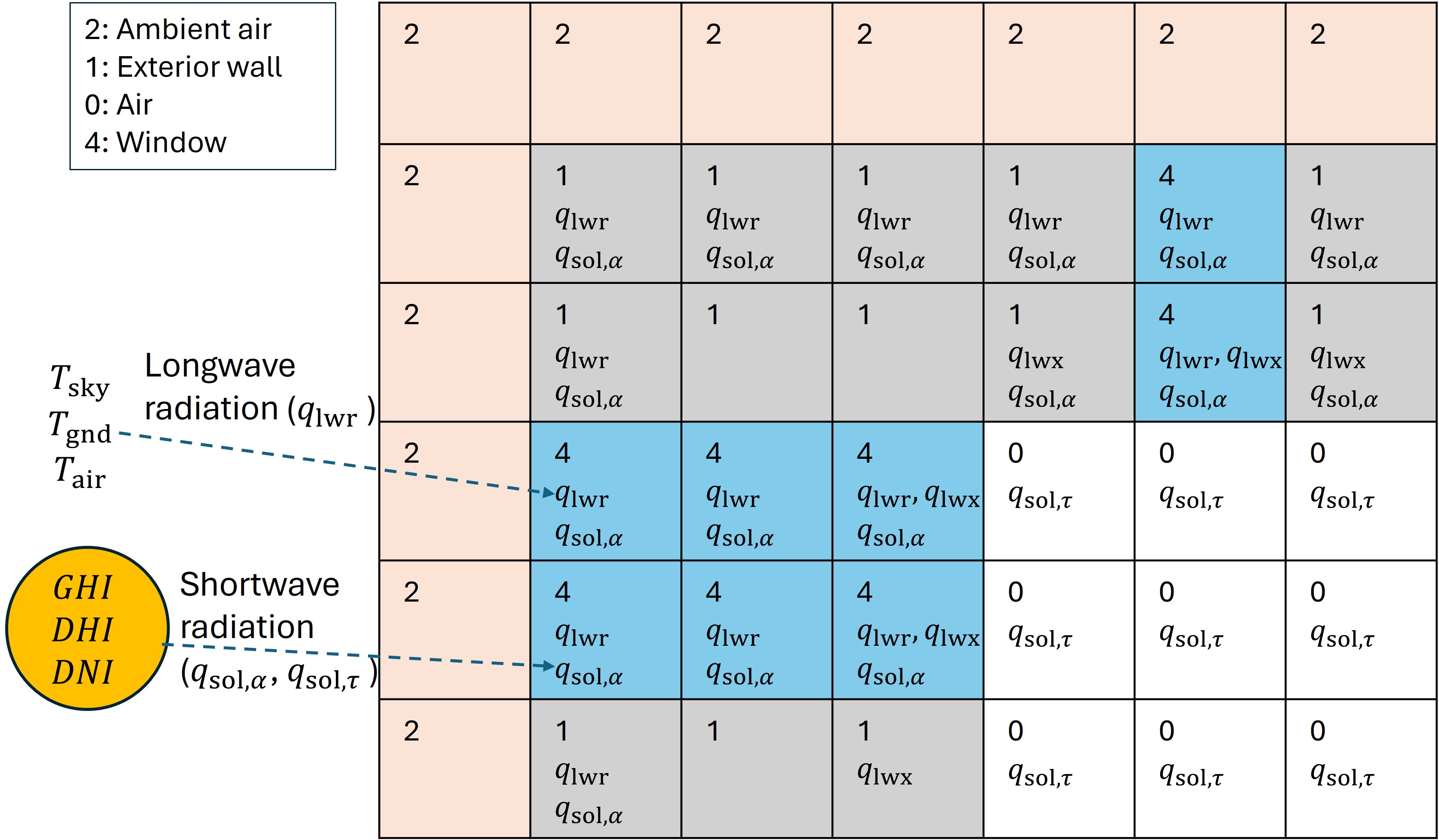}
  \caption{Assignment of radiative flux terms to CV types on a representative building cross-section. Exterior wall (type~1) and window (type~4) nodes receive exterior LW ($q_{\text{lwr}}$) and absorbed solar ($q_{\text{sol},\alpha}$) fluxes. Interior wall nodes adjacent to both exterior and interior air spaces additionally participate in interior LW exchange ($q_{\text{lwx}}$). Interior air nodes (type~0) receive transmitted solar flux ($q_{\text{sol},\tau}$), routed to mass nodes when enabled.}
  \label{fig:grid_radiation}
\end{figure}

Building on the interior radiation module described in \cite{Ham2025-ht}, this work augments the \texttt{sbsim} energy balance with three additional physical processes: exterior LW radiation, solar radiation, and interior thermal mass coupling. Figure~\ref{fig:grid_radiation} illustrates how these flux terms are assigned to the various CV types within the computational grid.

\subsection{Exterior Long-Wave Radiation}

Every exterior surface of a building exchanges LW radiation with the sky dome, the ground plane, and the surrounding atmosphere. Following the tilted-surface model commonly employed in building energy simulation \cite{US_Department_of_Energy2023-mh}, the net exterior LW flux on a surface with tilt angle $\phi$ from horizontal is:
\begin{equation}
\label{eq:qlwr}
q_{\text{lwr}} = \epsilon\sigma\bigl[F_{\text{gnd}}(T_{\text{gnd}}^4 - T_{\text{surf}}^4) + \beta F_{\text{sky}}(T_{\text{sky}}^4 - T_{\text{surf}}^4) + F_{\text{air}}(T_{\text{air}}^4 - T_{\text{surf}}^4)\bigr]
\end{equation}
where $\epsilon$ is surface emissivity, $\sigma$ the Stefan--Boltzmann constant, and $T_{\text{surf}}$ the exterior face temperature of the CV. The tilt-dependent view factors and sky correction parameter are defined as $F_{\text{gnd}}=\tfrac{1}{2}(1-\cos\phi)$, $F_{\text{sky}}=\tfrac{1}{2}(1+\cos\phi)$, $\beta=\sqrt{F_{\text{sky}}}$, and $F_{\text{air}}=F_{\text{sky}}(1-\beta)$. For vertical walls ($\phi=90^{\circ}$), $F_{\text{gnd}}=F_{\text{sky}}=0.5$.

To incorporate this flux into the tensorized framework, we construct an exterior LW radiation tensor $Q_{\text{lwr}}$ whose entries depend on the CV type. Corner CVs, which possess two exposed boundary faces, receive a contribution of $2\delta_x \cdot q_{\text{lwr}}$; edge CVs with a single exposed face receive $\delta_x \cdot q_{\text{lwr}}$; and interior envelope CVs receive $(\delta_x/k)\cdot q_{\text{lwr}}$. This scaling ensures dimensional consistency with the other terms in the energy balance after division by the common denominator.

\subsection{Solar Radiation}

Plane-of-array (POA) irradiance $G_{Ts}$ for each exterior surface orientation is computed from weather-file Global Horizontal Irradiance (GHI), Direct Normal Irradiance (DNI), and Diffuse Horizontal Irradiance (DHI) using the \texttt{pvlib} library \cite{F_Holmgren2018-fc}. This library handles solar position calculation, transposition of irradiance components to the tilted plane, and ground-reflected irradiance based on surface albedo, providing a well-validated and computationally efficient implementation.

Two distinct solar flux components arise from the computed POA irradiance. The absorbed solar flux $q_{\text{sol},\alpha}=\alpha\cdot G_{Ts}$ represents energy absorbed at the exterior face of opaque walls and glazing, where $\alpha$ is the solar absorptivity. The transmitted solar flux $q_{\text{sol},\tau}=\tau\cdot G_{Ts}$ represents energy that passes through transparent surfaces into the zone interior, where $\tau$ is the glazing transmissivity ($\tau=0$ for opaque surfaces). Corresponding tensors $Q_{\text{sol},\alpha}$ and $Q_{\text{sol},\tau}$ are constructed with the same CV-type-dependent scaling used for $Q_{\text{lwr}}$.

\subsection{Interior Thermal Mass}

\begin{figure}[htbp]
  \centering
  \includegraphics[width=0.6\linewidth]{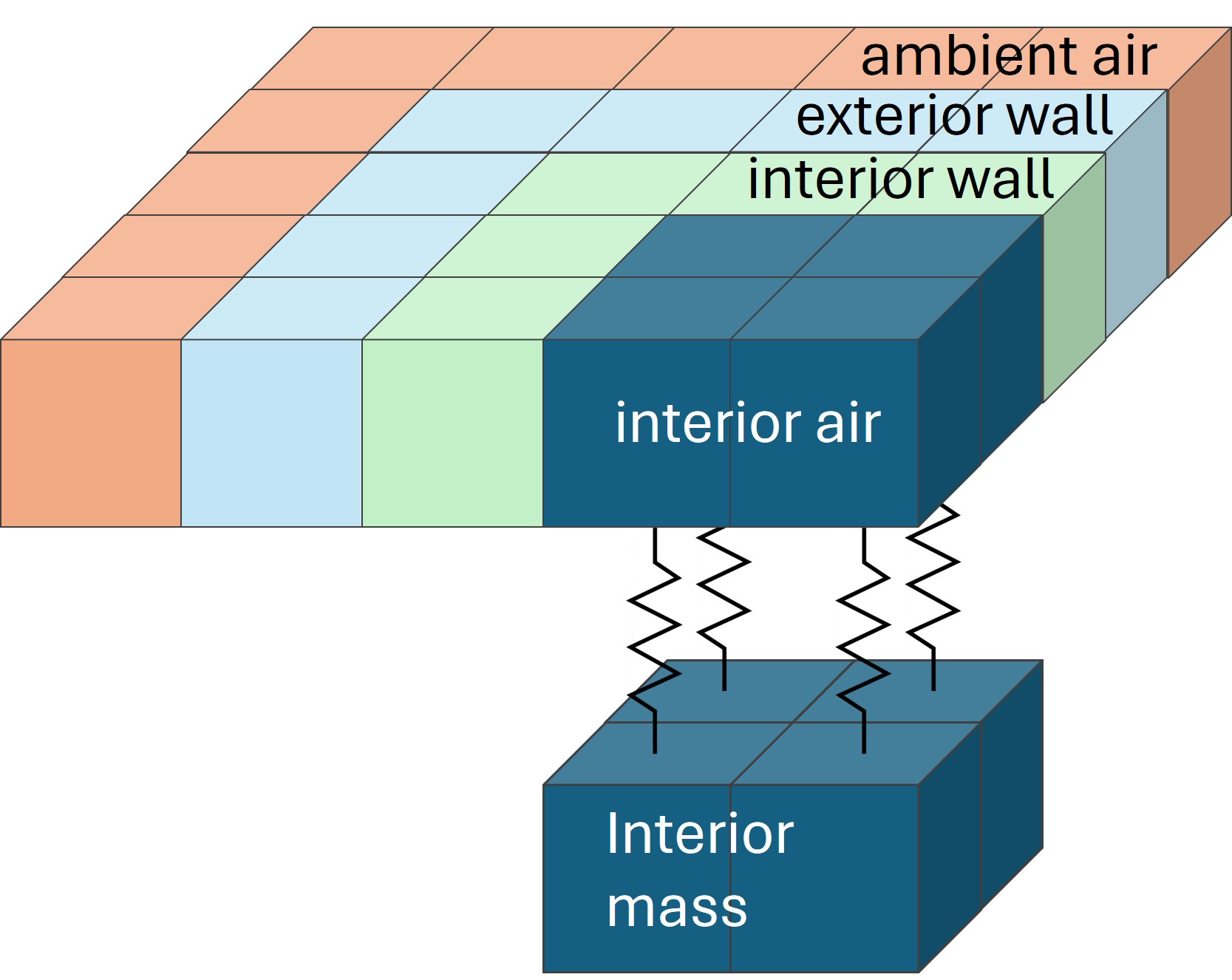}
  \caption{Interior mass node coupling. Each indoor air CV connects to a sub-grid mass node through a conductive resistance parameterized by $k_{\text{mass}}$ and $z$. Transmitted solar radiation $q_{\text{sol},\tau}$ is deposited directly onto mass nodes.}
  \label{fig:interior_mass}
\end{figure}

Real buildings contain furnishings, interior partitions, and floor slabs that absorb and store thermal energy, particularly the short-wave radiation transmitted through windows. To capture this effect, we introduce an interior mass temperature tensor $T_{\text{mass}}$ that is conductively coupled to each indoor air CV (Figure~\ref{fig:interior_mass}). The coupling uses a conductance $K_{\text{mass}}$ representing the effective thermal connection between the air volume and the mass layer, with the floor height $z$ serving as the characteristic length. When interior mass is enabled, the transmitted solar flux $Q_{\text{sol},\tau}$ is excluded from the air CV energy balance and instead deposited onto the mass nodes. After air temperatures converge at each timestep, the mass temperature updates explicitly as:
\begin{equation}
\label{eq:tmass}
T_{\text{mass}} = \frac{T + \dfrac{q_{\text{sol},\tau}\cdot z}{k_{\text{mass}}} + t_{0,\text{mass}}\odot T_{\text{mass}}^{(-)}}{1 + t_{0,\text{mass}}}
\end{equation}
where $t_{0,\text{mass}}=\rho_{\text{mass}}c_{\text{mass}}z^2/(k_{\text{mass}}\Delta t)$ is the mass temporal parameter, $\odot$ denotes element-wise multiplication, and $T_{\text{mass}}^{(-)}$ is the mass temperature from the preceding timestep.

\subsection{Complete Tensorized Energy Balance}

Combining all radiative and mass-coupling terms with the original conduction--convection formulation, the updated temperature equation becomes:
\begin{equation}
\label{eq:new_sbsim}
\begin{aligned}[t]
T ={} & \biggl[ Q_x + Vz\bigl(K_1 U^{-1}T_1 + H_1 T_\infty + K_3 U^{-1}T_3 + H_3 T_\infty\bigr) \\
      & + Uz\bigl(K_2 V^{-1}T_2 + H_2 T_\infty + K_4 V^{-1}T_4 + H_4 T_\infty\bigr) \\
      & + K_{\text{mass}}UVz^{-1}T_{\text{mass}} + Q_{\text{lwx}} + Q_{\text{lwr}} \\
      & + Q_{\text{sol},\alpha} + Q_{\text{sol},\tau} + \tfrac{C\rho UVz}{\Delta t}\,T^{(-)} \biggr] \\
      & \cdot \biggl[ Vz\bigl(K_1 U^{-1}{+}H_1{+}K_3 U^{-1}{+}H_3\bigr) \\
      & + Uz\bigl(K_2 V^{-1}{+}H_2{+}K_4 V^{-1}{+}H_4\bigr) \\
      & + K_{\text{mass}}UVz^{-1} + \tfrac{C\rho UVz}{\Delta t} \biggr]^{-1}
\end{aligned}
\end{equation}
where $Q_{\text{lwx}}$ is the interior LW exchange tensor computed via the pre-computed radiation matrix \cite{Ham2025-ht}, $Q_{\text{lwr}}$ is the exterior LW tensor from Eq.~\eqref{eq:qlwr}, $Q_{\text{sol},\alpha}$ and $Q_{\text{sol},\tau}$ are the absorbed and transmitted solar tensors, and $K_{\text{mass}}UVz^{-1}T_{\text{mass}}$ is the conductive coupling to interior mass. The denominator is augmented with the corresponding $K_{\text{mass}}UVz^{-1}$ term to maintain energy conservation. The iteration proceeds until the maximum CV temperature change falls below threshold $\epsilon$.

\section{Validation and Results}

We verified the tensorized formulation against an independently coded iterative finite-volume solver that applies energy balance equations node by node, sweeping sequentially until convergence. Both solvers were configured with an identical two-zone building floor plan on a $23\times12$ grid (276~CVs), a 300-second timestep, and boundary conditions including ambient, sky, and ground temperatures plus solar irradiance components (GHI, DNI, DHI) drawn from a representative weather dataset.

\begin{figure}[htbp]
  \centering
  \includegraphics[width=0.9\linewidth]{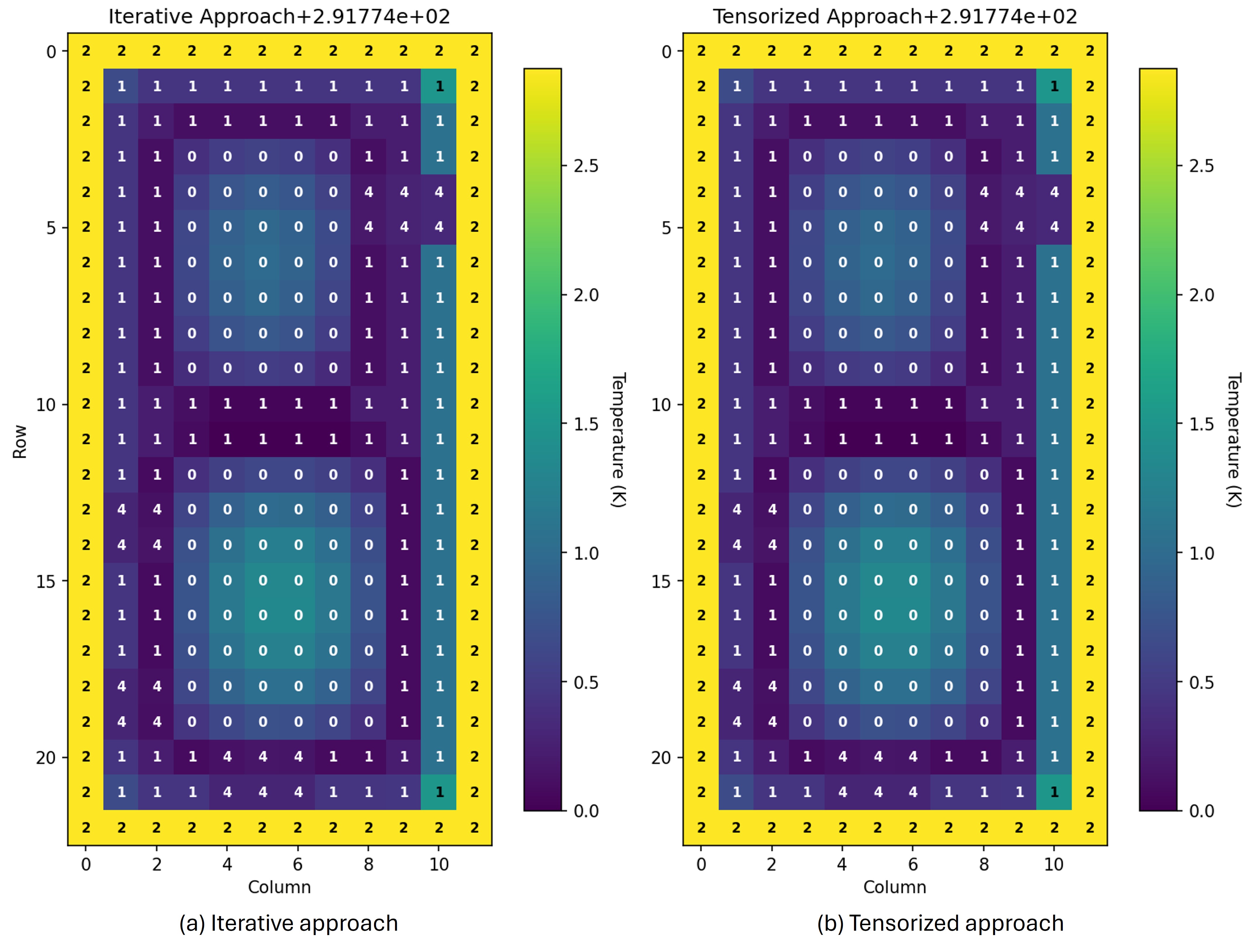}
  \caption{Temperature field from the iterative solver (left) and tensorized solver (right) with full radiative heat transfer (interior LWX + exterior LWR + solar). Cell annotations indicate CV type. Both methods produce numerically identical results. See Figure~\ref{fig:grid_radiation} for node type notation.}
  \label{fig:validation}
\end{figure}

Figure~\ref{fig:validation} compares the resulting temperature distributions with all radiative mechanisms enabled (interior LW exchange, exterior LW, and solar radiation). The two outputs are visually indistinguishable, and a pointwise comparison confirms agreement to at least five significant figures across all CVs. This level of agreement confirms that the tensorized formulation correctly implements the underlying physics encoded in the iterative energy balance.

Beyond numerical correctness, we evaluated computational performance by running both solvers for 10 simulation steps on the same hardware: an AMD Ryzen~7 5800X processor with 48\,GB DDR4 RAM and an NVIDIA RTX~3080 GPU, running Ubuntu~22.04 under WSL~2.0. Table~\ref{tab:benchmark} summarizes the timing results. The tensorized solver achieves a 4.19$\times$ speedup over the iterative baseline. Individual step times for the tensorized solver exhibit greater variance due to GPU scheduling and memory allocation patterns, but every step is substantially faster than its iterative counterpart. This advantage is expected to grow with larger building models because tensor operations scale more favorably on GPU hardware than sequential node-by-node iteration. For RL training workflows requiring millions of simulation episodes, this performance differential translates directly into reduced wall-clock training time.

\begin{table}[htbp]
\centering
\caption{Wall-clock execution times for 10 simulation steps on a $23\times12$ grid (276~CVs) with $\Delta t=300$\,s and convergence threshold $\epsilon=0.001$.}
\label{tab:benchmark}
\begin{tabular}{lcc}
\toprule
\textbf{Metric} & \textbf{Iterative} & \textbf{Tensorized} \\
\midrule
Total time (s) & 12.70 & 3.03 \\
Mean per step (s) & 1.27 & 0.30 \\
\midrule
\textbf{Speedup} & \multicolumn{2}{c}{\textbf{4.19$\times$}} \\
\bottomrule
\end{tabular}
\end{table}

\section{Conclusion and Future Work}

This paper presented a comprehensive radiative heat transfer implementation for the \texttt{sbsim} building energy simulator, extending prior work on interior LW exchange \cite{Ham2025-ht} with exterior LW radiation, short-wave solar radiation, and interior thermal mass coupling. All new physics modules are formulated as tensor operations and integrated into the existing tensorized finite-difference framework, preserving the computational efficiency that makes \texttt{sbsim} suitable for reinforcement learning applications.

We validated the complete radiative model by comparing the tensorized solver against an independently coded iterative finite-volume solver, demonstrating numerical agreement to five significant figures across all control volumes. Performance benchmarking showed that the tensorized approach delivers a 4.19$\times$ speedup over the iterative baseline on a 276-CV building grid, with greater advantages anticipated for larger geometries.

Future priorities include cross-validation against established tools such as EnergyPlus \cite{US_Department_of_Energy2023-mh} and the Modelica Buildings Library \cite{Wetter2011-dc}, expanding the HVAC equipment library beyond variable-air-volume systems to include heat pumps and radiant systems, and coupling the enhanced thermal model with an RL training pipeline to evaluate control performance improvements attributable to higher-fidelity radiation modeling.

\section*{Acknowledgments}
The authors would like to thank the developers of the original Smart Buildings Control Suite for making their work open-source and providing a strong foundation for this research. The simulator code and dataset are publicly available in the Google Open Source repository at \url{https://github.com/google/sbsim}.

\bibliographystyle{IEEEtran}
\bibliography{paperpile}

\end{document}